\documentclass[twoside]{article}
\usepackage{epsfig.sty}
\newcommand{\beq}{\begin{eqnarray}}
\newcommand{\eeq}{\end{eqnarray}}
\begin{document}
\title{$\pi+X$ Production via Transversly Polarized $p^{\uparrow}+p$ Collisions
and the Collins Fragmentation Function}
\author{Leonard S. Kisslinger\\
Department of Physics, Carnegie Mellon University, Pittsburgh, PA 15213\\
Ming X. Liu and Patrick McGaughey \\
P-25, Physics Division, Los Alamos National Laboratory, Los Alamos, NM 87545}
\date{}
\maketitle
\begin{abstract}
  This article is motivated by a Letter of Intent for a Drell-Yan experiment
to measure the sea quark Sivers function. We estimate $\pi^{+},\pi^{-}$ +X 
production, where X represents particles which are not detected, via proton 
collisions with a polarized proton target using the Collins fragmentation 
function derived using the parton distribution functions for polarized p-p 
collisions at E=$\sqrt{s}$=200 GeV.
\end{abstract}
PACS Indices:12.38.Aw,13.60.Le,14.40.Lb,14.40Nd
\vspace{1mm}

\section{Introduction}

  We estimate inclusive $\pi^{+},\pi^{-}$+X production, where X represents
particles not detected, via transversely polarized p-p collisions at 200 GeV, 
an extension of our recent work\cite{lsk15} on $D$ production via unpolarized 
p-p collisions  at E=$\sqrt{s}$=200 GeV. The E1039 Collaboration, see 
Ref\cite{LOI13-15} for the Letter of Intent, plans to carry out a Drell-Yan
experiment with a polarized proton target, with the main objective to
measure the Sivers function\cite{siv90}. A number of Deep Inelastic Scattering
experiments\cite{hermes09,compass09,jlab11} have measured non-zero values
for the Sivers Function. See these references for references to earlier 
experiments.

 Another important function is the Collins fragmentation function\cite{col93},
which describes the fragmentation of a transversly polarized quark into an
unpolarized hadron, such as a pion. For many years there have been theoretical 
studies of the Collins fragmentation functions, such as Refs\cite{atru97,bacc02,
ansel13}. Recently a study of extracting the Collins fragmentation function 
from experiment has been carried out\cite{kang15,kang16}. 

  Ref\cite{lsk15} used the method of Braaten et. al.\cite{bcfy95,bc96,fl95} to 
estimte the fragmentation of
a charm quark to a D, a meson consisting of a charm and anti-light quark.
In the present work we calculate  fragmentation of a polarized light
quark to a pion. Although both estimates of fragmentation use the 
method of Ref.\cite{bcfy95} the magnitudes of the resulting cross section
for $p^{\uparrow}+p \rightarrow \pi+X$ is much larger than $p+p \rightarrow 
D+X$, due to the large difference in the charm and light quark masses, as will 
be shown.

\section{Sivers and Collins Functions}

The Sivers and Collins functions are defined by the target assymmetry,
$A(\phi_h,\phi_S$), in the scattering of an unpolarized lepton beam by a
transversely polarized target\cite{jlab11}:
\beq
\label{sivcol}
     A(\phi_h,\phi_S)&\simeq&A_Csin(\phi_h+\phi_S)+
A_Ssin(\phi_h-\phi_S) \;,
\eeq
where $A_C,A_S$ are the Collins, Silvers moments with $\phi_h$  the hadron
momentum azimuthal angle and $\phi_S$ the target spin azimuthal angle with 
respect to the lepton scattering plane. 

In this section we briefly review the Sivers Function and the Collins
Fagmentation Function.

\subsection{Sivers Function}

See, e.g., Ref\cite{hermes09} for a discussion of the Sivers Function 
in terms of experimental cross sections.

  The Sivers term of the cross section for the production of hadrons using an 
unpolarized lepton beam on a transversely polarized target is\cite{hermes09}
\beq
\label{sivers}
 \sigma(\phi_h,\phi_S)&=& \sigma_{UU}|S_T|[2<sin(\phi_h-\phi_S)>_{UT}\times
sin(\phi_h-\phi_S) +...] \;,
\eeq
where $\phi_h$ and $\phi_S$ were defined above.  $\sigma_{UU}$ is the 
$\phi_h$-independent part of the polarization-independent cross section; and 
$UT$ denotes the unpolarized beam and transverse target polarization w.r.t. 
the virtual photon direction.  The Sivers function is obtained by an expansion
of $2<sin(\phi_h-\phi_S)>_{UT}$ in quarks using a quark-parton 
model\cite{hermes09,bm98}. As mentioned in the Introduction, a number of Deep 
Inelastic Scattering experiments have measured the Sivers function and 
obtained non-zero values.

\subsection{Collins Fragmentation Function}

The definition of the Collins Function\cite{kang15,kang16}is similar to that 
of the Sivers Function in Eq(\ref{sivers}):
\beq
\label{collins}
 \sigma_C(\phi_h,\phi_S)&\propto& F_{UU}(1+A_{UT}\times sin(\phi_h+\phi_S))
 \;,
\eeq
with $\phi_h$ and $\phi_S$ defined above, $F_{UU}$ the spin-averaged structure 
function, and $A_{UT}$ the asymmetry that can be 
calculated from quark distribution and fragmentation, discussed in detail in 
the next section. See Refs.\cite{kang15,kang16} for definitions of 
$\sigma_C,F_{UU}$ and derivation of $F_{UU}$.

From Ref\cite{bggm08} (also see Ref\cite{ggo03}  and earlier references for 
the derivation of $H^q \equiv H_1^{\perp q}$) the fragmentation probability to 
produce a hadron, $h$, from a transversely polarized quark in $e^+ e^-$ 
annihilation is given by the function
\beq
\label{polqfrag}
 D_{h q^{\uparrow}}(z,k_T^2)&=& D^q(z,K_T^2)+H^q(z,K_T^2)\frac{(\vec{k}\times
\vec{K}_T/k)\cdot s_q}{z M_h} \; ,
\eeq
where $M_h$ is the hadron mass, $\vec{k}$ the quark momentum, $s_q$  the quark
spin vector, $\vec{K}_T$ the hadron momentum transverse to $\vec{k}$, $z$ the 
light-cone momentum fraction of h wrt the fragmenting quark, and $D^q$ the 
unpolarized fragmentation function. The Collins fragmentation function is $H^q
\equiv H_1^{\perp q}$.
\newpage

\section{Differential $p^{\uparrow}+p \rightarrow \pi+X$ and
$p+p \rightarrow \pi+X$ cross sections}

This work on estimating the cross section $ \sigma_{p^{\uparrow}p\rightarrow \pi X}$
is an extension of our previous work on $J/\Psi,\Psi'(2S)$,
and $\Upsilon(nS)$ production\cite{klm11} for E=$\sqrt{s}$=200 GeV using
the color octet model\cite{cl96,bc96,fl95}, but for the production of pions
rather than heavy quark mesons. The differential cross section for a proton
collision with a proton polarized orthogonal to the scattering plane is
\beq
\label{Deltasig0}
      \frac{d\sigma_{p^{\uparrow}p\rightarrow \pi X}}{dy} &\simeq& 
  \frac{1}{x(y)}\Delta f_g(x(y),2m)\Delta f_g(a/x(y),2m) \sigma_{q^{\uparrow}+
g\rightarrow  \pi X}
\; ,
\eeq
where rapidity $y$ is defined in Eq(\ref{y(x)}), $\Delta f_g(x)$ is the gluon 
distribution function for polarized p-p collisions (see Ref\cite{klm11}) and 
$a=4m^2/s$, $m\simeq$ 3.5 MeV for light quarks; and with\cite{bcfy95}
\beq
\label{2}
 \sigma_{q^{\uparrow}+g\rightarrow \pi X}&=& 2 \sigma_{q+g\rightarrow q\bar{q}}
D_{q^{\uparrow} \rightarrow q \pi} \; ,
\eeq
where $\sigma_{q+g\rightarrow q\bar{q}}$ is similar to the charmonium production
cross section in Ref\cite{klm11}, but with light  quarks rather than charm
quarks. $D_{q^{\uparrow} \rightarrow q \pi}$ is the total fragmentation 
probability for a polarized light quark, $q$, to fragment to a $\pi(q\bar{q})$
meson plus a $q$ quark. 

  Note there are also contributions to $d\sigma_{p^{\uparrow}p\rightarrow 
\pi X}/dy$ from $\Delta f_d(x(y),2m) \Delta f_{\bar{d}}(a/x(y),2m)$
and $\Delta f_u(x(y),2m) \Delta f_{\bar{u}}(a/x(y),2m)$ quarks and 
anti-quarks. However, as shown in Ref\cite{klm11} these contributions are more 
than an order of magnitude smaller that those from gluons.

  The Collins fragmentation function  for a polarized light quark fragmenting
to a $q\bar{q}(\pi)$ meson plus $q$, a light quark, is illustrated in  
Figure 1 (see Ref\cite{bcfy95}, Sec.II, FIG 1). 
\begin{figure}[ht]
\begin{center}  
\epsfig{file=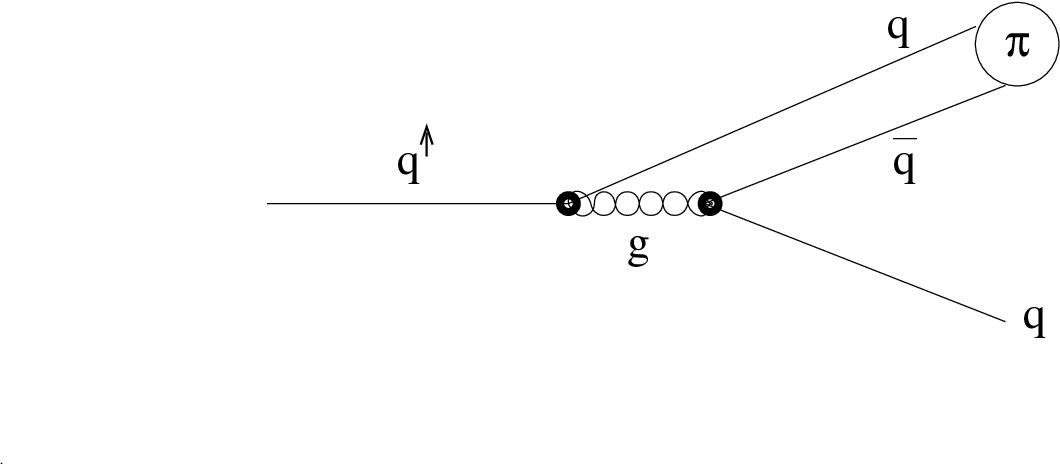,height=6 cm,width=12cm}
\caption{ Light quark fragmentation for $q^{\uparrow} \rightarrow q+q\bar{q}
 \rightarrow q+\pi $}
\label{}
\end{center}
\end{figure}

  Note that the target proton is polarized, leading the polarized quark
$q^{\uparrow}$. This also results in the cross section depending on the
gluon distribution functions $\Delta f_g(x)$ rather than  $f_g(x)$ used
in Ref.\cite{lsk15}
\newpage
Ref\cite{bcfy95} explains above Eq(31) that a 
$1/m_Q$ expansion was used so Eq(31) also gives light quark fragmentation,
$D_{q^{\uparrow} \rightarrow q\bar{q}}(z)$,  with a light quark polarized $q^{\uparrow}$ 
replacing a heavy unpolarized quark $Q$. Note also from Figure 1 that the $\pi$
production is non-perturbative as $g\rightarrow q\bar{q}\rightarrow \pi$ 
invoves a nonperturbative qluon-quark interaction. With $r \equiv 
(m_\pi-m_q)/m_\pi \simeq 1.0$, from Ref\cite{bcfy95} Eq(31) one finds
\beq 
\label{3}
 D_{q^{\uparrow} \rightarrow q\bar{q}}(z)&\simeq& N(6z+6z^2-15z^3-12z^4
+15z^5) \; ,
\eeq
with $0 \leq z \leq 1$ and

\beq 
\label{4}
     N &=&\frac{2 \alpha_s(\mu)^2 |R(0)|^2}{81 \pi m_q^3}\simeq 2.125 \times
 10^{-3}
\; ,
\eeq
where $\alpha_s(\mu)\simeq.26$ for $\mu\simeq m_q\simeq$ 3.5 MeV, and using for
$R(0)$, the radial wave function of the
$|q\bar{q}>$ meson, $|R(0)|^2\simeq 4 m_q^3$.
From Eqs(\ref{3},\ref{4}),
\beq
\label{5}
     D_{q^{\uparrow} \rightarrow q\bar{q}}&=&\int_{0}^{1} dz
D_{q^{\uparrow} \rightarrow q\bar{q}}(z)=2.87 \times 10^{-3} 
\eeq

In order to extract $H^q$ from Eq(\ref{polqfrag}) one also needs to
evaluate the unpolarized fragmentation function, $D^q$. From 
From Ref\cite{bcfy95}, Eq(33) with r=$m_q/(m_q+m_Q) \simeq 0$,
\beq
\label{Dunpolarized}
        D^q&=& \int_{0}^{1}dz D^q(z)=3N\frac{8}{15} \simeq 3.4 \times 10^{-3}
\eeq

A fit to the parton distribution functions for polarized p-p collisions
for Q$\simeq$ 1 GeV obtained from CTEQ6\cite{CTEQ6} in the x range needed for 
the present work at $\sqrt{s}$=200 GeV is
\beq
\label{6}
\Delta f_g(x) &\simeq & 15.99-700.34 x+13885.4 x^2-97888. x^3 \; .
\eeq

  From Eqs(\ref{Deltasig0},\ref{2})
\beq
\label{7}
 \frac{d\sigma_{p^{\uparrow}p\rightarrow \pi X}}{dy}&=& Aqq \Delta f_g(x(y),2m)
\Delta f_g(a/x(y),2m) \frac{dx(y)}{dy} \frac{1}{x(y)} 
D_{q^{\uparrow} \rightarrow q\bar{q}} \; ,
\eeq
with rapidity $y$
\beq
\label{y(x)}
      y &=& \frac{1}{2} ln (\frac{E + p_z}{E-p_z}) \nonumber \\
        x(y) &=& 0.5 \left[\frac{m}{E}(\exp{y}-\exp{(-y)})+\sqrt{(\frac{m}{E}
(\exp{y}-\exp{(-y)}))^2 +4a}\right] \; , 
\eeq
where $Aqq$ is the matrix element for $q\bar{q}$ production\cite{klm11} 
modified by an effective mass $m_s$. For E=200 GeV $Aqq= 7.9 \times 10^{-4} 
(1.5 GeV/m_s)^3 nb$. With $m_s=3.5 MeV$, $Aqq=0.60 \times 10^{-5}$ nb.
\newpage
 
 For unpolarized $ p p\rightarrow \pi X$ the differential cross section is
\beq
\label{unpol}
 \frac{d\sigma_{p p\rightarrow \pi X}}{dy}&\simeq& Aqq  f_g(x(y),2m)
 f_g(a/x(y),2m) \frac{dx(y)}{dy} \frac{1}{x(y)} D_q \; ,
\eeq
with $D_q$ given in Eq(\ref{Dunpolarized}) and
\beq
\label{fg}
 f_g(y) &=& 1334.21 - 67056.5 x(y) + 887962.0 x(y)^2
\eeq

 From Eqs(\ref{6},\ref{7},\ref{5},\ref{y(x)}) we find 
$\frac{d\sigma_{p^{\uparrow}p\rightarrow \pi X}}{dy}$, and from Eqs(\ref{unpol},
\ref{fg},\ref{Dunpolarized}) we find $\frac{d\sigma_{pp\rightarrow \pi X}}{dy}$
shown in Figure 2.

\vspace{1cm}

\begin{figure}[ht]
\begin{center}
\epsfig{file=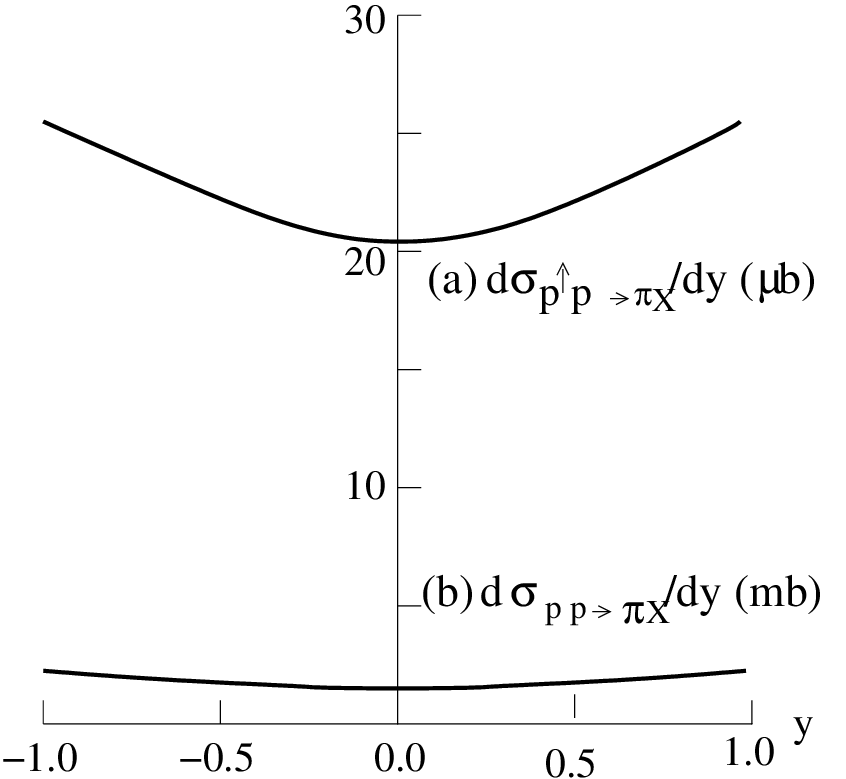,height=10cm,width=12cm}
\end{center}
\caption{(a) $d\sigma/dy$ ($\mu$b) for E=200 GeV polarized p-p collisions 
producing $\pi+X$; (b) $d\sigma/dy$ (mb) for E=200 GeV unpolarized p-p 
collisions producing $\pi+X$} 
\end{figure}
\newpage
\section{Conclusions}

We have estimated the production of pions via polarized $p^{\uparrow}-p$
collisions, deriving the Collins fragmentation function needed for the
$p^{\uparrow}-p \rightarrow \pi+X$ cross section, and for comparison the
unpolarized $p-p \rightarrow \pi+X$ cross section. Although the Drell-Yan 
unpolarized experiment described in Ref.\cite{LOI13-15} can determine the 
Sivers but not the Collins function, future experiments with $p^{\uparrow}-p$ 
collisions might be able to measure the cross section for $\pi+X$ production. 
From $\frac{d\sigma_{p^{\uparrow}p\rightarrow \pi X}}{dy}$, using 
Eqs(\ref{polqfrag},\ref{7},\ref{6},\ref{Dunpolarized}), one would be able to 
measure $H^q \equiv H_1^{\perp q}$, the Collins fragmentation function. The 
results of the present article could thereby be tested.

\vspace{1cm}
\Large{{\bf Acknowledgements}}\\
\normalsize
This work was supported in part by the DOE contracts W-7405-ENG-36 and 
DE-FG02-97ER41014. The author LSK was a visitor at Los Alamos National 
Laboratory, Group P25. The authors thank Dr. Xiaodong Jiang,LANL Group P25
and Dr. Zhong-Bo Kang, LANL Group T-2, for helpful discussions. The authors
declare that there is no conflict of interest regarding the publication of 
this paper.

\end{document}